==Preprint v.1==

## Origin of pressure-induced band gap tuning in crystalline lactose

*Igor A. Fedorov*

Kemerovo State University, 650000, Kemerovo, Russia

E-mail: ifedorov@kemsu.ru (I. A. Fedorov)

**Abstract**

Lactose is widely used in agri-food and pharma industries. New materials that can be obtained from renewable sources are currently being sought. I have studied the effect of hydrostatic pressure on structural properties of currently known forms of crystalline lactose within the framework of density functional theory with van der Waals interactions. The computed parameters have good agreement with experimental data. The effect of mechanical deformations on electron structure of crystalline lactose was also studied. Compression of the crystal leads to the band gap increase. The analysis of partial density of states of lactose crystals was performed. The band gap of different forms of crystalline lactose was also computed using the quasiparticle $G_0W_0$ approximation.

**Keywords:** density functional theory, many-body perturbation theory, molecular crystals, lactose

### Introduction

Organic molecular crystals have been studied over many years. A long-standing interest in the crystals has been maintained due to their unusual properties, which can find practical applications [1–4]. Molecules in molecular crystals are held together by van der Waals forces, and some molecules can also be held together by hydrogen bonds. Atoms inside a molecule are bound together by covalent and ionic bonds. Thus, the molecule has a rigid structure and is weakly deformed under external impact. These properties are revealed in high anisotropy of the mechanical properties of molecular crystals. This can lead to susssch an interesting effect as negative linear compressibility in some cases [5,6].

Currently, there is growing interest in organic systems. It is due to search of new as environmentally-friendly as possible materials ("green chemistry"). Petroleum-based raw materials are widely used in production, while rare earth elements are used for electron devices production. Therefore, scientists are currently searching for materials that are easily processed and synthesized from renewable raw materials. Amino acids and sugars are considered as a source of these raw materials. Moreover amino acids, as an example, often have several chiralities, which makes them a promising candidate for optical properties. Organic electronics (OLED, AMOLED, OFET), organic laser and organic biosensor is currently widely used [7,8]. However, the synthesis of materials that can be used in electronic devices presents some difficulties. For example, pentacene and its derivatives, which are widely used in organic diodes, are obtained using the chemical vapor deposition (CVD) method. This is due to the fact that aromatic compounds are poorly soluble in liquids and unstable in melts. However, research aimed at improving methods for synthesizing organic crystals continues [9]. The advantage of sugars is the fact that their crystalline forms can be obtained from solutions. Thus, it makes it easier and cheaper to obtain large crystals necessary for practical production.

Lactose has two anomers (α-lactose and β-lactose), which differ in the configuration of the terminal hydroxyl group of the glucose residue (Fig. 1). Lactose crystals can be formed from molecules of various shapes. Atoms in lactose molecule form covalent bonds. Molecules in crystalline lactose (Fig.1) are held together due to van der Waals forces as well as due to hydrogen bonds. Several forms of crystalline lactose are currently known (Table 1) and fully characterized crystallographically [10–14]. Three crystalline forms have been characterized for the α-anomer [13]: the α-lactose monohydrate (α-L-H₂O), hygroscopic anhydrous α-lactose (α-LH), and stable anhydrous α-lactose (α-LS). The stable anhydrous form of α-lactose has been obtained by the dehydration of α-lactose monohydrate in methanol [12]. The β-anomer has only one crystalline form (β-L) [13,15]; mixed-lactose compounds have also been identified with different stoichiometries (αβ-LH) [14,16–18]. The crystal structures of the L-H₂O form have also been determined [11,19–21]. The structure of the α-LH [10] and α-LS [12] forms have been determined *ab initio* from powder X-ray diffraction experiments.

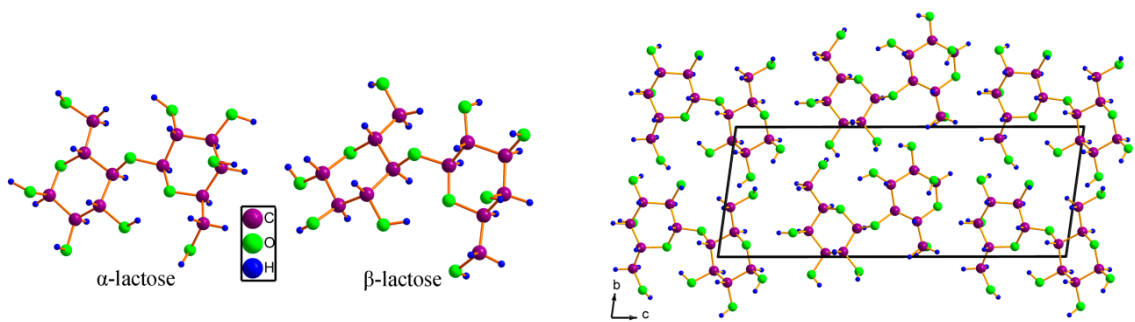

Figure 1. Molecules from crystalline α-lactose and β-lactose. The unit cell of crystalline αβ-lactose (αβ-LH$_T$).

Table 1. Crystalline forms of lactose.

| Form | Abbreviation | SpGrp | Z | T (K) | Refcode | Type | Reference |
|---|---|---|---|---|---|---|---|
| α-lactose (hygroscopic) | α-LH | $P2_1$ | 2 | 293 | EYOCUQ | PXRD | Platteau et al. [10] |
| α-lactose monohydrate | α-L-H₂O | $P2_1$ | 2 | 150 | LACTOS11 | SX | Smith et al.[11] |
| α-lactose (stable anhydrous) | α-LS | $P1$ | 2 | 293 | EYOCUQ01 | PXRD | Platteau et al [12] |
| β-lactose | β-L | $P2_1$ | 2 | 100 | BLACTO02 | SX | Garnier et al [13] |
| αβ-lactose | αβ-LH$_T$ | $P1$ | 2 | 120 | LAKKEO01 | SX | Guiry et al.[16] |
| αβ-lactose | αβ–LH$_M$ | $P2_1$ | 4 | 100 | LAKKEO02 | PXRD | Nicholls et al.[14] |
| Notes: SpGrp – space group; refcode – Cambridge Structural Database (CSD [22]) refcode; SX - structure determined by single-crystal X-ray diffraction; PXRD – structure determined by powder X-ray diffraction; T – triclinic, M – monoclinic. | | | | | | | |

It is interesting to study the effect of mechanical deformations on electron structure of crystalline lactose. Pressure brings molecules closer to each other, which contributes to better understanding of the properties of crystals under consideration. Molecules can noticeably approach each other even in the low-pressure region along directions where van der Waals forces play a key role. Fedorov *et al* [3] investigated the electronic and elastic properties of α- and β-lactose (α-LH and β-L) crystals. The purpose of this study is to investigate the electronic properties of all known forms of crystalline lactose under pressure.

**Computational details**

A plane-wave pseudopotential approach within density functional theory was used to compute properties of crystalline lactose. The Quantum ESPRESSO (QE)[23] with the functional of Perdew, Burke, and Ernzerhof (PBE)[24], was used to perform the calculations. The ultrasoft pseudopotentials of the Rabe-Rape-Kaxiras-Joannopoulos type were used for calculations[25]. The crystal structures were calculated within the Broyden-Fletcher-Goldfarb-Shanno (BFGS) scheme [26]. The energy cutoff equals 65 Ry. The **k**-point grid[27] are 3×2×3 (α-LH), 3×1×2 (α-L-$H_2O$), 2×1×3 (α-LS), 3×2×2 (β-L), 3×2×1 (αβ-$LH_T$) and 3×1×2 (αβ–$LH_M$).

Geometry relaxation was completed when all components of all forces are smaller than 0.1 mRy (a.u.)$^{-1}$. DFT-D3(BJ) scheme was used to take into account van der Waals interaction[28–31]. The band gaps of the crystals were calculated using the CRYSTAL17[32,33], using the PBE0 hybrid functional[34] and standard 6-31G* basis set[35]. Also, I have used the implementation of the $G_0W_0$ approximation [36,37] provided by the code YAMBO[38,39]. I used Troullier-Martins norm-conserving pseudopotentials [40]. The cutoff energy of plane waves was set to 95 Ry. The dielectric function is calculated using the plasmon-pole approximation. The GW dielectric matrix cutoff is 12 Ry. To obtain convergence, I used 300 unoccupied states.

**Results and discussion**

**Elastic properties**

Table 2 presents the computed and experimental values of lattice parameters of crystalline lactose. Theoretical values are in good agreement with experimental data. These results confirm that the DFT-D3(BJ) scheme correctly describes van der Waals interaction between molecules in crystalline lactose. Thus, it is possible to study mechanical and electronic properties of lactose crystals using DFT-D3(BJ).

The effect of hydrostatic pressure on structural properties of α- and β-lactose has been studied in earlier works [3]. Tables S1–S6 present the computed values of the lattice parameters of the six forms of crystalline lactose under pressure. Figure 2 and 3 presents the pressure effect on the lattice parameters and volume lactose crystals. Hydrostatic pressure decreases lattice parameters of crystalline lactose. All crystals have anisotropy of mechanical properties, which is due to hydrogen bond between lactose molecules. Thus, it is more difficult to compress a crystal along hydrogen bond than along the direction where van der Waals forces play a key role.

Table 2. The lactose computed and experimental lattice parameters at zero pressure.

| Form | Method | $a$ (Å) | $b$ (Å) | $c$ (Å) | $\alpha$ (°) | $\beta$ (°) | $\gamma$ (°) | $V$ (Å³) |
|---|---|---|---|---|---|---|---|---|
| α-LH | theory[3] | 7.7303 | 19.6373 | 4.8451 | 90.000 | 104.137 | 90.000 | 713.214 |
| | exp[10] | 7.7795 | 19.6931 | 4.9064 | 90.000 | 103.6909 | 90.000 | 730.32 |
| α-L-H$_2$O | theory | 4.7231 | 21.5207 | 7.7586 | 90.000 | 106.641 | 90.000 | 755.592 |
| | exp[11] | 4.7830 | 21.540 | 7.7599 | 90.000 | 105.911 | 90.000 | 768.85 |
| α-LS | theory | 7.6109 | 19.7204 | 4.8735 | 92.660 | 106.016 | 96.496 | 696.242 |
| | exp[12] | 7.6522 | 19.864 | 4.9877 | 92.028 | 106.261 | 97.153 | 720.18 |
| β-L | theory[3] | 4.9111 | 13.1532 | 10.8166 | 90.000 | 92.935 | 90.000 | 697.805 |
| | exp[13] | 4.9325 | 13.2700 | 10.7792 | 90.000 | 91.554 | 90.000 | 705.29 |
| αβ-LH$_T$ | theory | 5.0673 | 7.4247 | 19.4497 | 84.305 | 87.196 | 74.643 | 701.955 |
| | exp[16] | 5.030 | 7.593 | 19.374 | 81.026 | 85.044 | 74.247 | 702.7 |
| αβ-LH$_M$ | theory | 4.9275 | 38.5109 | 7.5709 | 90.000 | 105.983 | 90.000 | 1381.122 |
| | exp[14] | 5.0044 | 38.6364 | 7.6007 | 90.000 | 106.200 | 90.000 | 1411.26 |

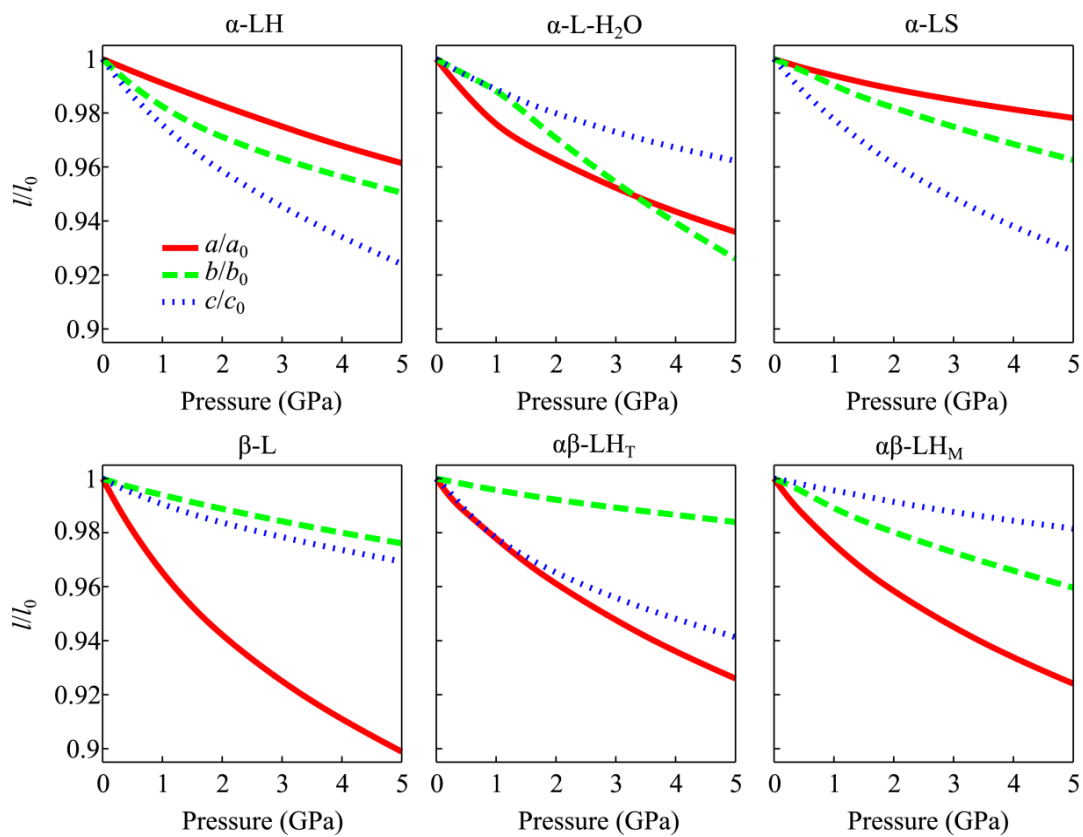

Figure 2. Computed pressure dependencies of the relative lattice parameters $a/a_0$, $b/b_0$,

and $c/c_0$ for lactose crystals.

The β-lactose crystal is compressed by 10% along the $a$-axis, and by 2% along the $b$-axis. Note the almost identical change in volume under pressure (Fig. 2) for α-LS and αβ–LH$_M$, as well as α-LH and αβ-LH$_T$. At the same time, the lattice parameters of these crystals change differently. In addition, these crystals have different symmetries of unit cells.

Tables S7–S12 presents the complete elastic constants of lactose crystals. These constants allowed us to calculate the basic mechanical characteristics of the studied crystals (Table 3). The triclinic αβ-lactose crystal has the maximum anisotropy. The minimum anisotropy is observed in α-lactose monohydrate (α-L-H2O). In this crystal, water molecules form hydrogen bonds with lactose molecules, which is manifested in the minimum value of βmax among the studied crystals. Thus, the presence of water molecules between lactose molecules leads to the fact that the crystal is more difficult to compress. In general, the basic characteristics of lactose crystals have different values. Thus, the packing of molecules is manifested in the mechanical properties of crystalline lactose.

Table 3. Isotropic aggregate elastic properties based on the Voigt-Reuss-Hill averages. The bulk modulus (*B*), Young's modulus (*E*), shear modulus (*G*), Poisson ratio (μ), Pugh ratio (*G/B*), minimum and maximum values for crystals of lactose are computed by using DFT-D3(BJ).

| Crystal | *B*, GPa | *E*, GPa | *G*, GPa | μ | *G/B* | $\beta_{min}$, $GPa^{-1}$ | $\beta_{max}$, $GPa^{-1}$ | $\dfrac{\beta_{max}}{\beta_{min}}$ |
|---|---|---|---|---|---|---|---|---|
| α-LH [3] | 24.04 | 32.13 | 12.58 | 0.28 | 0.52 | 0.005 | 0.033 | 6.31 |
| α-L-$H_2O$ | 23.34 | 29.80 | 11.57 | 0.29 | 0.50 | 0.007 | 0.027 | 3.86 |
| α-LS | 23.39 | 30.47 | 11.87 | 0.28 | 0.51 | 0.006 | 0.031 | 5.17 |
| β-L [3] | 19.11 | 30.43 | 12.33 | 0.23 | 0.64 | 0.008 | 0.042 | 5.18 |
| αβ-$LH_T$ | 21.26 | 25.15 | 9.65 | 0.30 | 0.45 | 0.005 | 0.039 | 7.80 |
| αβ-$LH_M$ | 25.16 | 32.23 | 12.53 | 0.29 | 0.50 | 0.005 | 0.031 | 6.20 |

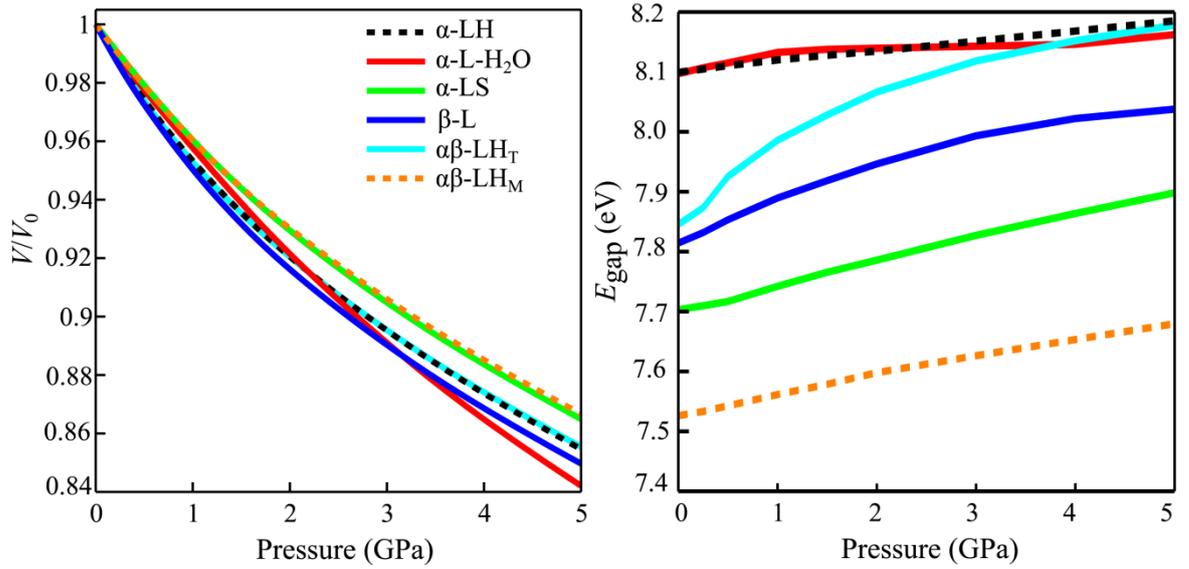

Figure 3. Computed pressure dependencies of the relative volume ($V/V_0$) and energy gap ($E_{gap}$) for lactose crystals.

**Band gaps**

Figures 4 and S1-S5 present band structures of lactose crystals. The labeled ***k***-points and band paths correspond to the SeeK-path project (http://www.materialscloud.org/tools/seekpath/) [41,42]. Band structures have weak dispersion, which is typical for molecular crystals. This is due to weak intermolecular interaction. The energy gap values are presented in Table 4. The band gap for all the forms of crystalline lactose is above 7.5 eV. Thus, it was established that all studied lactose crystals are dielectrics.

Table 4. Calculated energy gaps (eV) lactose crystals.

| Method | PBE0 | | $G_0W_0$ | |
|---|---|---|---|---|
| $P$ (GPa) | 0 | 5 | 0 | 5 |
| α-LH | 8.10 | 8.18 | 8.70 | 8.65 |
| α-L-$H_2O$ | 8.10 | 8.16 | 8.91 | 8.99 |
| α-LS | 7.70 | 7.90 | 8.34 | 8.45 |
| β-L | 7.81 | 8.04 | 8.40 | 8.63 |
| αβ-LH$_T$ | 7.84 | 8.18 | 8.57 | 8.73 |
| αβ-LH$_M$ | 7.53 | 7.68 | - | - |

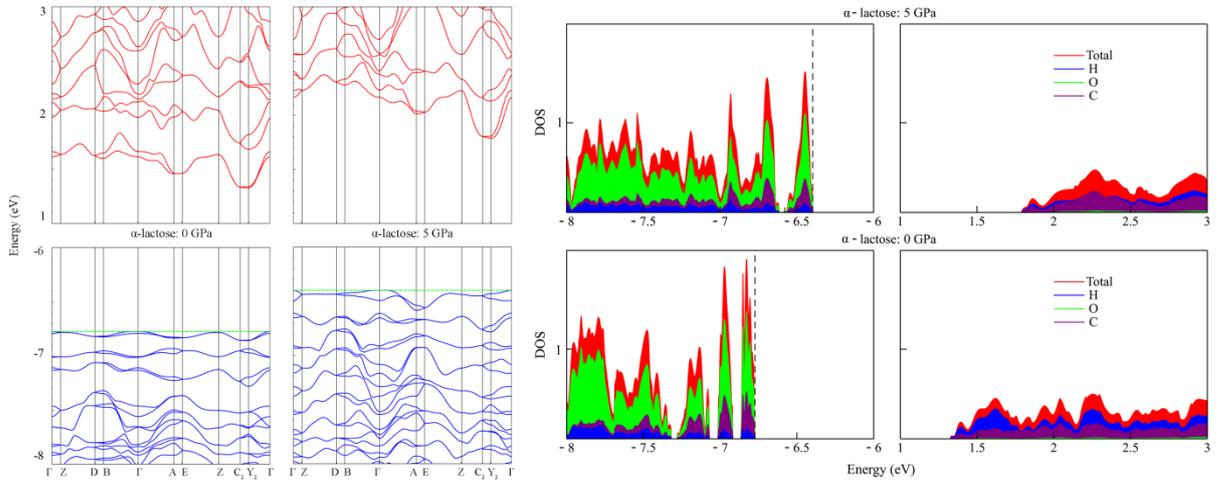

Figure 4. Band structure $E(\mathbf{k})$, DOS and PDOS of crystalline α-lactose within PBE0 approximation at ambient and hydrostatic pressure of 5 GPa. The top of the valence band indicated by the Fermi level (dashed line).

Figure 3 shows the dependencies of band gaps on pressure ($E_{gap}(P)$) for different forms of crystalline lactose. It is established that the band gap slightly changes under hydrostatic pressure. At the same time hydrostatic pressure leads to volume changes of lactose crystals by ~13-16% respectively. It has also been established that the band gap of the lactose crystals under consideration increases. Such effect is rarely seen and deserves to be studied. So, additional computations of the band gap for some forms of crystalline lactose have been performed using the quasiparticle method ($G_0W_0$). The calculations within quasiparticle approximation confirmed that the band gap increases under hydrostatic pressure. However, the energy gap of the α-LH crystal calculated within the $G_0W_0$ framework decreases slightly under pressure.

Figures 4 and S1-S5 shows DOS and PDOS for lactose crystals under consideration. As can be seen from PDOS analysis, the main contribution to formation of the top of valence band is made by *p*-orbitals of oxygen atoms. The bottom of conduction band is mainly formed by *p*-orbitals of carbon atoms, while the contribution of oxygen atoms is very small. Significant contribution is made by *s*-orbitals of hydrogen. Thus, different response of crystalline orbitals to hydrostatic pressure leads to slight increase in energy gap under hydrostatic pressure.

Let us consider the effect of compression on the band structure of the α-lactose crystal within the framework of DFT-PBE0. The response of top of valence band and bottom of conduction band of crystalline α-lactose to hydrostatic pressure is studied. Dispersion of the top of valence band is 0.08 eV and 0.16 eV at 0 and 5 GPa. Dispersion of the bottom of conduction band at the same pressure is 0.4 and 0.57 eV. The pressure of 5 GPa leads to increase of the value of top of valence band by 0.38 eV and bottom of conducting band by 0.46 eV. As a result, the band gap increases by 0.08 eV. Similar behavior is observed for the energy gaps for other forms of lactose crystals.

The energy gaps for uniaxial compression are presented in Table 5 and S13-16. In most cases, the band gap increases under uniaxial compression along the crystallographic axes. Compression along the *b*-axis of the α-LH crystal leads to a value of $E_{gap}$=7.86 eV. The band gap is 8.00 eV when compressed along the *a*-axis. This value is 0.10 eV greater than the value of the band gap at 5 GPa. This behavior is not observed for other forms of lactose. The band gap for the β-L crystal under compression along the *c* axis is almost the same as the band gap at 5 GPa, equal to 8.17 and 8.18 eV, respectively.

Table 5. The band gaps (eV) for α-lactose and β-lactose at hydrostatic and uniaxial compression, which is close to the value of 7.90 eV at 5 GPa.

| | α-LH | | | | β-L | | | |
|---|---|---|---|---|---|---|---|---|
| Press | $a$ (Å) | $b$ (Å) | $c$ (Å) | $E_{gap}$ | $a$ (Å) | $b$ (Å) | $c$ (Å) | $E_{gap}$ |
| 0 GPa | 4.9111 | 13.1532 | 10.8166 | 7.70 | 7.7303 | 19.6373 | 4.8451 | 8.10 |
| 5 GPa | 4.4146 | 12.8382 | 10.4844 | 7.90 | 7.4322 | 18.6646 | 4.4770 | 8.18 |
| compres, $a$ | 4.4146 | 13.1532 | 10.8166 | 8.00 | 7.4322 | 19.6373 | 4.8451 | 8.14 |
| compres, $b$ | 4.9111 | 12.8382 | 10.8166 | 7.86 | 7.7303 | 18.665 | 4.8451 | 8.12 |
| compres, $c$ | 4.9111 | 13.1532 | 10.4844 | 7.72 | 7.7303 | 19.6373 | 4.4770 | 8.17 |

* compres – uniaxial compression along the crystallographic axis

In the case of α-LS and αβ-$LH_T$ crystals, uniaxial compression along the crystallographic axes does not allow obtaining band gap values close to those at 5 GPa (Table S14-15). This is explained by the fact that the unit cells of these crystals have triclinic symmetry *P1*. It has the minimal symmetry among Fedorov groups. It is the only lattice type that itself has no symmetry other than the identity operation (E). Thus, hydrostatic pressure changes the values of the three angles that describe the crystal lattices, which gives more options for the arrangement of molecules relative to each other. Therefore, the energy gaps during uniaxial compression of α-LS and αβ-$LH_T$ crystals differ more strongly from the values corresponding to hydrostatic pressure.

It is well known that decrease in the volume of crystal within free electron model leads to increase in the dispersion of energy bands, as well as decrease in the band gap. The band gap decreases as the crystal is compressed, bringing the atoms closer together, resulting in increased overlap of atomic orbitals and vice versa. This mechanism is also carried out for lactose crystals, which consist of molecules. In this case, the compression of the crystal leads to the convergence of the molecules and the reduction of the length of the hydrogen bonds between the ones. At the same time, some covalent bonds (O-H) within the molecules increase slightly, resulting in a decrease in the overlap of atomic orbitals. As a result of a combination of several effects, the band gap of crystalline lactose changes only slightly under pressure.

### Conclusions

The pressure effect on structural properties of several forms of crystalline lactose has been studied using first principle calculations (DFT-D3(BJ)). It has established that bulk compression modulus for different forms is within the range of ~19 to 25 GPa. This value is typical for molecular crystals with hydrogen bonds between molecules. Besides, hydrogen and van der Waals bonds bond is also manifested in high anisotropy of mechanical properties of crystalline lactose.

Having studied the band structures, the energy gap is established to increase under pressure. The band gap (DFT-PBE0) increases by 0.06 and 0.34 eV for α-L-$H_2O$ and αβ-$LH_T$, respectively. These values are the minimum and maximum values for the forms of crystalline lactose under consideration. It is established that the bottom of conduction band is formed by *p*-orbitals of carbon atoms, while the top of valence band is formed *p*-orbitals of oxygen and carbon atoms. Different response of crystalline orbitals to mechanical compression leads to increase in energy gap. It has been established that the width of the forbidden gap of crystals with a triclinic unit cell changes more than crystals with a monoclinic unit cell. Small change in the band gap under mechanical deformations may be of interest for practical applications.

### Data availability statement

Data will be made available on request.

### Acknowledgments

The author gratefully acknowledge the Center for collective use "High Performance Parallel Computing" of the Kemerovo State University for providing the computational facilities. The author acknowledge support from the Ministry of Science and Higher Education of the Russian Federation (project no. FZSR-2024–0005).

### ORCID iD

Igor Fedorov https://orcid.org/0000-0002-6209-9830

### Declaration of Competing Interest

The author declares that he has no known competing financial interests or personal relationships that could have appeared to influence the work reported in this paper.